# Propagating phonons coupled to an artificial atom

Martin V. Gustafsson[1,2,*], Thomas Aref[1], Anton Frisk Kockum[1],
Maria K. Ekström[1], Göran Johansson[1], and Per Delsing[1,†]

[1] Microtechnology and Nanoscience, Chalmers University of Technology, Kemivägen 9, SE-41296, Göteborg, Sweden.

[2] Department of Chemistry, Columbia University, NWC Building, 550 w. 120th street, New York, NY, 10027, USA.

* Correspondence to: mg3465@columbia.edu

† Correspondence to: per.delsing@chalmers.se

Quantum information can be stored in micromechanical resonators, encoded as quanta of vibration known as phonons. The vibrational motion is then restricted to the stationary eigenmodes of the resonator, which thus serves as local storage for phonons. In contrast, we couple propagating phonons to an artificial atom in the quantum regime, and reproduce findings from quantum optics with sound taking over the role of light. Our results highlight the similarities between phonons and photons, but also point to new opportunities arising from the unique features of quantum mechanical sound. The low propagation speed of phonons should enable new dynamic schemes for processing quantum information, and the short wavelength allows regimes of atomic physics to be explored which cannot be reached in photonic systems.

The quantum nature of light is revealed and explored in its interaction with atoms, which can be either elemental or artificial. Artificial atoms typically have transition frequencies in the microwave range and can be designed on a microchip with parameters tailored to fit specific requirements. This makes them well suited as tools to investigate fundamental phenomena of atomic physics and quantum optics. In the form of superconducting qubits, they have seen extensive use in closed spaces (electromagnetic cavities), where they have ample time to interact with confined microwave radiation (*1–3*). These experiments have recently been extended to quantum optics in open one-dimensional (1D) transmission lines, where the atom interacts with itinerant microwave photons (*4–7*). We present an acoustic equivalent of such a system, where the quantum properties of sound are explored, rather than those of light.

At the intersection between quantum informatics and micromechanics, recent milestones include the coupling between a superconducting qubit and a vibrational mode (*8, 9*), hybrids of mechanical resonators and electrical microwave cavities (*10*), and the use of mechanics to interface between microwaves and optical photons (*11, 12*). The system we present here is another manifestation of mechanics in the quantum regime, but one that differs fundamentally from the suspended resonators mentioned above. In our case, the phonons are not bound to the eigenmodes of any structure, but consist of Surface Acoustic Waves (SAWs) which propagate freely over long distances, before and after interacting with an atom in their path.

In the domain of quantum information, SAWs with high power have been used to transport electrons and holes in semiconductors (*13–15*). This stands in contrast with our use of SAWs, where the power is much too low to transport charge carriers, and we instead focus on the quantum nature of the phonons themselves.

We do this by coupling an artificial atom directly to the SAWs via piezoelectricity, so that this mode of interaction becomes the dominant one for the atom. This means that we can communicate with the atom bidirectionally through the SAW channel, exciting it acoustically as well as listening to its emission of propagating surface phonons.

The idea that this might be feasible was put forward in previous work (*16*), where the use of a single-electron transistor as a sensitive probe for SAWs was demonstrated. Earlier work on the interaction between phonons and two-level systems include the demonstration of SAW absorption by quantum dots (*17*) and theoretical treatments of phonon quantum networks (*18*) and phononic coupling to dopants in silicon (*19, 20*).

**The acoustically coupled atom**. Although there are several types of SAWs, we use the term to denote Rayleigh waves (*21–23*), which propagate elastically on the surface of a solid within a depth of approximately one wavelength. At and above radio frequencies (RF), the SAW wavelength is short enough that the surface of a microchip can serve as a medium of propagation. By use of a piezoelectric substrate, SAWs can be generated efficiently from electrical signals, and converted back to the electrical domain after propagating acoustically over a long distance on the chip. This is used extensively in commercial applications such as microwave delay lines and filters (*22–24*).

The primary component in a microelectronic SAW device is the interdigital transducer (IDT), which converts power from the electrical to the acoustic domain and vice versa. In its simplest form it consists of two electrodes, each made of many long fingers deposited as thin films on a piezoelectric substrate. The fingers of the two electrodes are interdigitated so that an AC voltage applied between the electrodes produces an oscillating strain wave in the surface of the substrate. This wave is periodic in both space and time, and radiates as a SAW away from each finger. The periodicity $\lambda_{IDT}$ of the fingers defines the acoustic resonance of the IDT, with the frequency given by $\omega_{IDT} = 2\pi v_0 / \lambda_{IDT}$ where $v_0$ is the SAW propagation speed. When the IDT is driven electrically at $\omega = \omega_{IDT}$, the SAWs emanating from all fingers add coherently, resulting in strong acoustic beams launching from the IDT in the two directions perpendicular to the fingers.

Our sample is fabricated on the (100) surface of a semi-insulating GaAs substrate, chosen for its piezoelectric and mechanical properties (Fig. 1A). The IDT is visible on the left side of the sample with a zoom-in in Fig. 1B. It has $N_{IDT} = 125$ finger pairs, with an overlapping width of $W = 25\mu m$. The fingers, made of aluminum capped with palladium, are aligned so that the SAW propagates in the [011] direction of the crystal, at a speed $v_0 \approx 2900$ m/s.

In our case, the IDT makes use of internal reflections to achieve strong electro-acoustic power conversion which would otherwise be infeasible (*22, 23*). This results in a narrow bandwidth of 1 MHz around $\omega_{IDT} / 2\pi = 4.8066$ GHz. The IDT is coupled to a low-noise cryogenic HEMT amplifier via a circulator and an isolator. Through the circulator and the IDT, we can launch a SAW beam toward the artificial atom, which is shown to the right in Fig. 1A with zoom-ins in Figs. 1D and 1E. Conversely, the IDT can pick up leftward-propagating SAW phonons emitted or reflected by the atom. All experiments were done in a dilution refrigerator with a base temperature of 20 mK, that is, with $k_B T \ll \hbar \omega_{IDT}$. At this low temperature, the charge carriers in the substrate are fully frozen out and the Bose-Einstein distribution gives a population of less than $10^{-4}$ thermal phonons per mode around $\omega_{IDT}$.

The artificial atom in our setup is a superconducting qubit of the transmon type (*25*), positioned 100 μm away from the IDT (Figs. 1A and D-E). A transmon consists of a Superconducting Quantum Interference Device (SQUID) shunted by a large geometric capacitance $C_{tr}$ with charging energy $E_C = e^2/(2C_{tr})$. The Josephson energy $E_J$ of the SQUID can be tuned with a magnetic flux $\Phi$. The Josephson inductance $L_J = \hbar^2/(4e^2 E_J)$ forms a resonant circuit together with $C_{tr}$, and the nonlinearity of $L_J$ gives rise to the anharmonic energy spectrum that is characteristic for an atom. The transmon is ideally suited for coupling to SAWs since the shunt capacitance can be designed as a finger structure, like an IDT. The charge on $C_{tr}$ then relates directly to the mechanical strain of Rayleigh waves in the underlying substrate surface.

The periodicity of the capacitor fingers defines the resonance frequency where the acoustic coupling of the qubit is strongest. By design, the qubit and the IDT have the same acoustic resonance frequency, $\omega_{IDT}$. The finger structure of the qubit has $N_{tr} = 20$ periods. In contrast with the IDT, each period of the qubit consists of four fingers, in a configuration that greatly diminishes internal mechanical reflections (*22, 23, 26*). This, along with the lower $N_{tr}$ compared with $N_{IDT}$, means that the qubit has a much wider acoustic bandwidth than the IDT (~250 MHz). From the geometry and materials of the device, we estimate $C_{tr} = 85$ fF. In addition to the strong coupling to SAWs, the qubit couples weakly to an RF gate through a capacitance $C_{gate}$. The gate (in contrast with the IDT) has a high bandwidth, so we can use it to excite qubit transitions away from $\omega_{IDT}$ as well as to apply RF pulses.

The acoustic coupling rate of the qubit, $\Gamma_{10}$, is an important characteristic of the system, with $1/\Gamma_{10}$ representing the average time it takes the qubit to relax from the first excited state $|1\rangle$ to the ground state $|0\rangle$ by emitting an acoustic phonon at the transition frequency $\omega_{10}$. The acousto-electric conversion of a finger structure is commonly represented as a complex and frequency-dependent acoustic admittance element $Y_a$, where electrical dissipation represents conversion to SAWs (*22, 23*). By inserting this element into a semiclassical model of a transmon, we get the circuit shown in Fig. 1F. Here, the qubit is approximated as a harmonic oscillator, and the model is thus not valid for qubit states beyond the first excited one, $|1\rangle$. When $L_J$ is adjusted so that the transition frequency $\omega_{10}$ between $|0\rangle$ and $|1\rangle$ resonates with $\omega_{IDT}$, $Y_{a,tr}$ is real-valued and we get the coupling strength as the power loss rate of the parallel LCR circuit. Using this model, we find $\Gamma_{10} = c_g^2 N_{tr} K^2 \omega_{10} / \sqrt{2} = 2\pi \times 30$ MHz, where $c_g \approx 0.8$ is a geometry factor and

$K^2 = 0.07\%$ is a material parameter that defines the strength of the piezoelectric coupling [(*27*), semi-classical model].

To analyze our system quantitatively, we have developed an extended model which takes the anharmonic nature of the qubit into account [(*27*), full quantum model] (*28–33*). A fully quantum mechanical model for the transmon (*25*) gives its energy levels as
$E_n \approx -E_J + \sqrt{8E_J E_C}(n+\frac{1}{2}) - \frac{E_C}{12}(6n^2 + 6n + 3)$. In our extended model we also need to account for the spatial extention of the qubit, since it interacts with SAWs over a distance of $N_{tr}$ wavelengths. We do this by considering one interaction point per finger and accounting for the SAW phase shifts between the different points. However, we assume that the propagation time along the qubit is short compared with the inverse coupling frequency, $v_0/(N_{tr}\lambda) \gg \Gamma_{10}$. This means that each emitted phonon leaves the qubit entirely before the next one is emitted, an assumption that is valid in our experiments. It is interesting to note that the opposite limit can be reached, where dressed states should form between excitations in the qubit and phonons localized within its finger structure. Our model predicts the same nonlinear reflection for phonons that has previously been observed for photons (*4*, *5*), and we compare it with experimental data in Figs. 2, 3, and 5.

**Acoustic reflection measurements**. An atom in an open 1D geometry reflects weak resonant coherent radiation perfectly (in the absence of pure dephasing). This is also predicted both by our semi-classical and quantum mechanical models. However, at irradiation powers comparable to or larger than one photon per relaxation time, the $|1\rangle$ state of the qubit becomes populated to an appreciable degree, which reduces its ability to reflect phonons of frequency $\omega_{10}$ and leads to an increase in transmission (*4*, *5*, *7*, *34*).

We measure the nonlinear acoustic reflection of the qubit by applying a coherent microwave tone of frequency $\omega$ to the IDT, which transmits part of the power in the form of a SAW propagating toward the qubit.

On IDT resonance ($\omega = \omega_{IDT}$), approximately 25% of the applied electrical power reflects against the IDT without converting to acoustic power (Fig. 2A). Here the qubit is tuned off resonance so that $\omega_{10} \gg \omega$. Of the power that leaves the IDT in the form of SAWs, half is emitted in the rightward direction. With all losses accounted for, approximately 8% of the applied electrical power reaches the qubit in the form of a SAW [(*27*), extracted sample parameters]. Of the acoustic power that reflects against the qubit, an appreciable part converts back to electrical power in the IDT and can be detected.

The qubit resonance frequency $\omega_{10}$ is modulated by the magnetic flux $\Phi$ applied through the SQUID loop, with a periodicity $\Phi_0 = h/2e$. As we tune the flux, we observe an increase in the reflected SAW when $\omega_{10}$ coincides with $\omega_{IDT}$. As shown in Figs. 2B and 2C, we can fit the flux modulation $\omega_{10}(\Phi/\Phi_0)$ to these resonance points. The absence of reflection outside the frequency band of the IDT shows that acoustic coupling strongly dominates over any electrical crosstalk between the IDT and the qubit.

A key characteristic of reflection against an atom in one dimension is the nonlinear dependence of the reflection coefficient on the power of the applied coherent tone: Only when the flux of incoming phonons is much lower than one per interaction time, $P_{SAW}/\hbar\omega \ll \Gamma_{10}$, does the qubit produce full reflection. As the power increases, the $|1\rangle$ state of the qubit becomes partly populated, which reduces its ability to reflect phonons of frequency $\omega_{10}$. This power-dependent saturation is shown in Fig. 2F. By fitting our theoretical model to these data, we find good agreement for $\Gamma_{10}/2\pi = 38\,\text{MHz}$ with negligible dephasing, which implies full reflection from the qubit in the low-power limit. All theoretical plots use the same values for the fitted parameters [(*27*), extracted sample parameters].

**Electrical driving in the steady state**. In addition to the acoustic excitation discussed above, we can address transitions in the qubit with the electrical gate and use the IDT to pick up the SAW phonons that the qubit emits. We expect the qubit to selectively emit one-phonon states when driven electrically on resonance, in a process similar to the nonlinear reflection observed under acoustic driving (Fig. 3A). With the frequency of the RF signal applied to the gate fixed at $\omega_{IDT}$, we observe the dependence of the emitted phonon flux on the qubit detuning $\Delta\omega_{10} = \omega_{10} - \omega_{IDT}$ as well as the applied RF power $P_{gate}$. As $P_{gate}$ increases from zero, we first see acoustic emission from the qubit at $\Delta\omega_{10} = 0$. At higher power, additional peaks show up for $\Delta\omega_{10} > 0$, which correspond to the excitation of higher states of the qubit by multiple photons, and subsequent relaxation into multiple phonons. The offsets between the peaks reflect the anharmonicity of the qubit. Fitting the positions of these multiphonon peaks allows us to determine the Josephson energy $E_{J0} = E_J(\Phi = 0)$ and $E_C$ of the qubit to good accuracy. We find $E_{J0}/h = 22.2\,\text{GHz}$ and $E_C/h = 0.22\,\text{GHz}$. This charging energy agrees well with our geometry-based estimate of the transmon capacitance.

**Time domain experiments**. The slow propagation of SAWs compared with electromagnetic waves allows us to clearly establish that the qubit couples to the IDT via phonons rather than photons. We do this by applying microwave pulses to the gate and studying the signal that reaches the IDT in the time domain. Fig. 4 shows the results of such experiments using 1μs long pulses at frequency $\omega_{IDT}$. When the qubit is tuned far away from resonance ($\Delta\omega_{10} \gg \Gamma_{10}$), we see a crosstalk signal reaching the IDT from the gate. This is virtually independent of frequency, and we attribute it to stray capacitance between the electrical transmission lines.

Also when the qubit is tuned close to its resonance, $\Delta\omega_{10} \approx 0$, the crosstalk signal is the first to rise above the noise floor. This leading edge serves as a time reference, showing the arrival of the electrical pulse to the gate. After the crosstalk edge, it takes another ~40ns before we observe the emission from the qubit, which corresponds to the acoustic propagation time from the qubit to the IDT. This shows unequivocally that the signal from the qubit is phononic. The phase of the SAW phonons emitted coherently by the qubit is sensitive to $\Delta\omega_{10}$, and with small variations around $\Delta\omega_{10}$, we can go from the case where the qubit emission and the crosstalk add constructively (blue) to the case where they add destructively (black).

Since the IDT partly reflects phonons impinging on it from the qubit, additional features can be seen which correspond to acoustic round trips from the IDT to the qubit and back. These acoustic reflections add gradually with time, superimposed on the ring-up of the IDT, until the steady-state signal is established after approximately 1μs. Just as in the steady-state experiments, the acoustic emission relative to the applied gate power increases with decreasing power $P_{gate}$, until it saturates at $P_{gate}/\hbar\omega \lesssim \Gamma_{10}$.

**Hybrid two-tone spectroscopy**. To characterize the dynamics of the phonon-qubit interaction in greater detail, we perform hybrid two-tone spectroscopy, where a continuous acoustic probe tone with low power is launched from the IDT toward the qubit at fixed frequency $\omega_{IDT}$ and its reflection is measured. At the same time, we vary the qubit detuning $\Delta\omega_{10}$ and apply a continuous electrical control tone to the gate, with varying frequency $\omega_{gate}$ and power $P_{gate}$. When the qubit absorbs photons from the gate, we observe an impact on its reflection of phonons from the IDT (Fig. 5).

For low values of $P_{gate}$, the acoustic reflection is modulated only by the qubit detuning, as demonstrated in Figs. 2B-E. For higher $P_{gate}$ and when $\omega_{gate}$ coincides with $\omega_{10}$, the control tone contributes to the population of the $|1\rangle$ state of the qubit. When the frequencies of the probe and control tones coincide, this results in saturation of the $|0\rangle \to |1\rangle$ transition, as seen in Fig. 2F. If the control tone is tuned to populate the |1⟩ state of the qubit and the condition $\omega_{21} = \omega_{IDT} = 2\pi \times 4.8066\,\text{GHz}$ is fulfilled, the $|1\rangle \to |2\rangle$ transition exhibits nonlinear acoustic reflection.

For still higher control powers, we observe a rich set of spectral features, which agrees well with our model [(*27*), full quantum model]. Of particular interest is the Autler-Townes doublet, which is caused by Rabi splitting of the $|1\rangle$ state at strong electrical driving of the $|1\rangle \to |2\rangle$ transition (*35, 36*).

**Outlook and summary**. Our SAW device occupies a middle ground between fixed mechanical resonators and transmission lines for free photons, and it is relevant to compare its features with both of these related systems.

Although their itinerant nature is an essential property of SAW phonons, they can also be confined into cavities by on-chip Bragg mirrors. Such cavities compare well with suspended resonators also in other respects than their natural integration with propagating waves: They can be fabricated for mode energies well above $k_B T$ for temperatures attainable with standard cryogenic equipment, and since the motion takes place directly in the cooled substrate, thermalization is excellent. The few available studies also indicate that high-frequency SAW cavities can have comparatively high quality factors at low temperature (*37, 38*).

In comparison to photons, SAW phonons have several striking features. Their speed of propagation is around $10^5$ times lower, and their wavelength at a given frequency correspondingly shorter. The slow speed means that qubits can be tuned much faster than SAWs traverse inter-qubit distances on a chip. This enables new dynamic schemes for trapping and processing quanta.

SAW phonons furthermore give access to a regime where the size of an atom substantially exceeds the wavelength of the quanta it interacts with. This is the opposite of the point-like interaction realized so far in photonics, cavity QED and circuit QED (*28*). In our device, the qubit is a modest factor $N_{tr} = 20$ times longer than the SAW wavelength, but this can be extended substantially in a device designed for such investigations.

In the device presented here, the coupling strength between SAWs and the qubit is also moderate. This is necessary to discriminate between the qubit transition energies due to the low anharmonicity of the transmon design. However, in a strongly piezoelectric material such as LiNbO$_3$, the many coupling points of an IDT-shaped qubit should make it possible to reach the regime of "ultrastrong coupling" (*31, 39, 40*), $\Gamma_{10} \gtrsim \omega_{10}/10$ and even "deep strong coupling" (*41*), $\Gamma_{10} \gtrsim \omega_{10}$, which are difficult to access with the standard electrical dipole coupling of photonic systems.

In conclusion, we have demonstrated non-classical interaction between surface acoustic waves and an artificial atom in the regime of strong coupling. Our experiments suggest that phonons can serve as propagating carriers of quantum information, in analogy with itinerant photons in quantum optics. The data are in good quantitative agreement with a theoretical quantum model, which captures the nonlinear acoustic reflection and relaxation of the atom. These results open up new possibilities in quantum experiments, since SAWs can reach regimes of coupling strength and time domain control that are not feasible with photons.

**Acknowledgments:** We acknowledge financial support from the Swedish research council (VR), the Wallenberg foundation, and the EU through the ERC and the SCALEQIT project. The samples were made at the Nanofabrication Laboratory at Chalmers. We acknowledge fruitful discussions with Peter Leek, Riccardo Manenti, and Daniel Niesner. M.V.G. acknowledges funding from the Wenner-Gren Foundations and support from Xiaoyang Zhu and Philip Kim during the manuscript preparation.

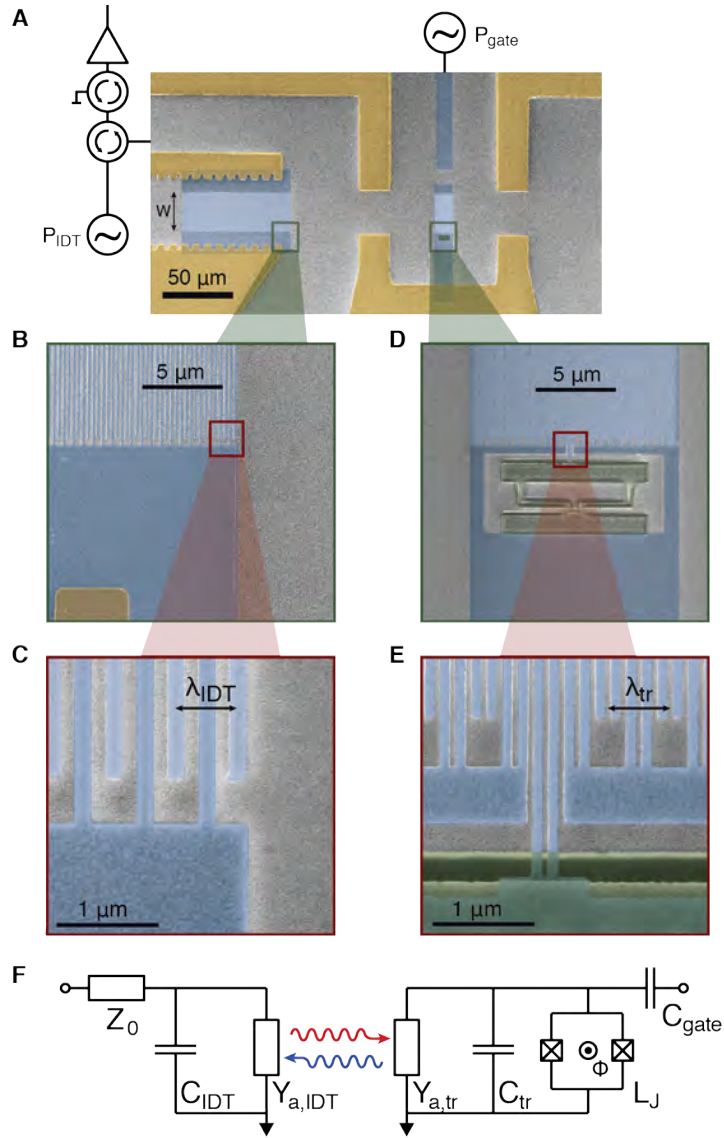

**Fig. 1**. Sample and experimental setup. **(A)** Electron micrograph of the sample (top view in false color). The interdigital transducer (IDT), shown to the left, converts electrical signals to surface acoustic waves (SAWs) and vice versa. It has $N_{IDT} = 125$ periods of fingers, each consisting of one finger from each electrode, with a periodicity of 600nm. The width of the SAW beam is given by the overlap of the fingers, W = 25μm. SAWs from the IDT propagate a distance of 100μm before reaching the qubit, which is shown to the right. All electrodes without explicit connections are grounded. **(B-C)** Zoom-ins on the IDT. **(D-E)** Zoom-ins on the transmon qubit. The qubit has $N_{tr} = 20$ finger periods, with double fingers to suppress internal mechanical reflections (*22, 23, 26*). **(F)** Semi-classical circuit model for the qubit. $C_{tr}$ is the geometric capacitance of the finger structure. It is shunted by a SQUID, which acts as a non-linear inductance $L_J$ that can be adjusted with a magnetic flux $\Phi$. $Y_{a,tr}$ is the acoustic admittance

element which can pick up SAWs from the IDT (red arrow) to produce electrical excitation in the qubit, and re-generate them with a phase shift (blue arrow) [(27), semi-classical and full quantum model]. In addition to SAWs from the IDT, radio-frequency signals applied through the gate capacitance $C_{gate}$ can be used to address transitions in the qubit.

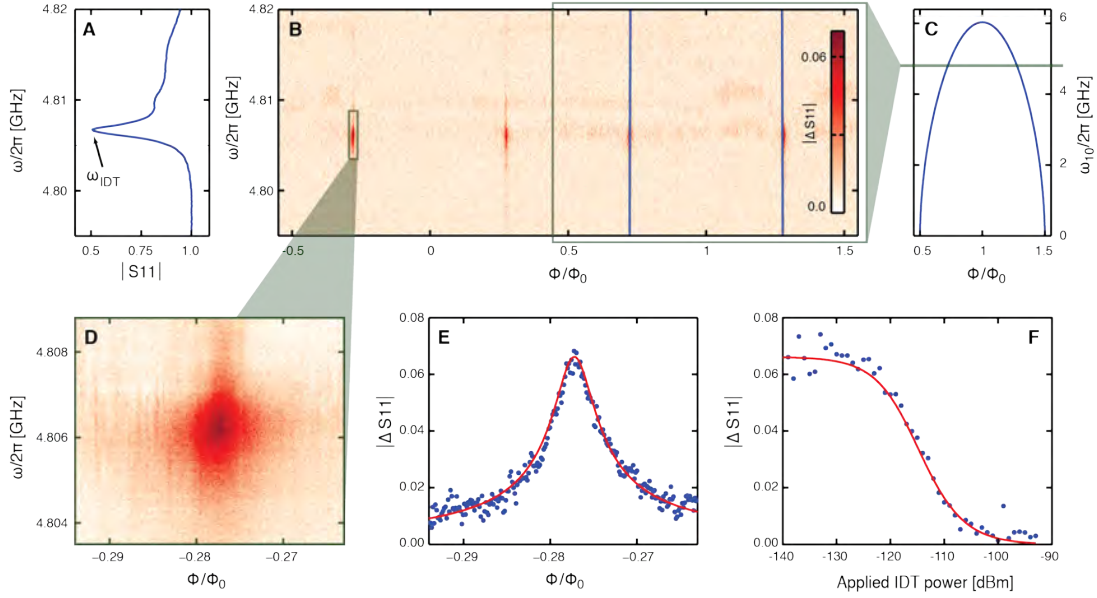

**Fig. 2**. Acoustic reflection measurements. **(A)** Reflection coefficient |S11| of the IDT *vs* frequency $\omega$, with the qubit far detuned ($\omega_{10} \gg \omega$). The dip in the curve represents conversion of applied electrical power into SAWs. The complex background reflection acquired with the qubit far detuned has been subtracted from panels B, D, E, and F, which thus show the change $|\Delta S11|$ in total reflection coefficient caused by the qubit. **(B)** $|\Delta S11|$ *vs* frequency and qubit flux, $\Phi/\Phi_0$. As $\Phi/\Phi_0$ increases, the first transition frequency of the qubit, $\omega_{10}$, periodically tunes in and out of the slim band where the IDT can address it acoustically. On resonance, the qubit reflects the incoming SAW beam. **(C)** Periodic modulation of $\omega_{10}$ with magnetic flux, fitted to the reflection peaks in panel B. **(D)** Zoom-in on one reflection peak from panel B. **(E)** Cross section through panel D at $\omega = \omega_{IDT}$ (blue). A fit to the theoretical expression for the qubit reflection (red) gives the acoustic coupling rate, $\Gamma_{10}/2\pi = 38\,\text{MHz}$. **(F)** Flux modulation of $|\Delta S11|$ *vs* power applied to the IDT. The rate at which the qubit reflects phonons from the incoming coherent beam is limited by the coupling rate $\Gamma_{10}$. Thus, the acoustic reflection coefficient of the qubit goes to zero for $P_{IDT}/\hbar\omega_{IDT} \gg \Gamma_{10}$, and the flux modulation vanishes. This saturation at the single-phonon level is evidence for the two-level nature of the qubit.

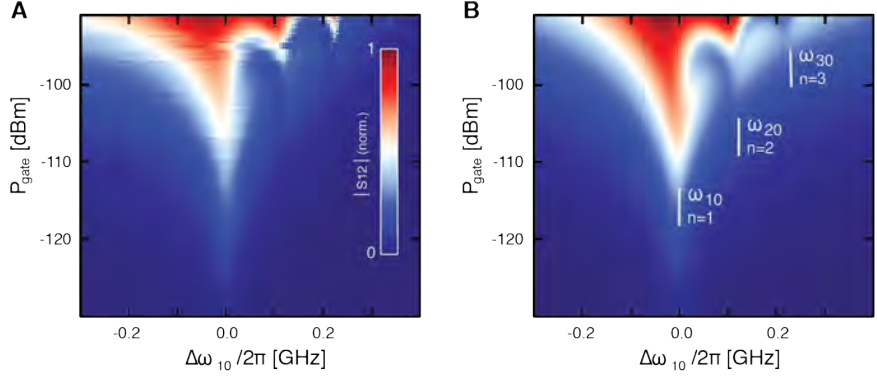

**Fig. 3**. Driving the qubit from the gate and listening to its acoustic emission with the IDT. **(A)** Transduction coefficient $|S12|$ from the gate to the IDT *vs* driving power and qubit detuning. We apply a continuous RF signal to the gate at fixed frequency, $\omega = \omega_{IDT}$, and adjust the detuning of the first qubit transition from this frequency, $\Delta\omega_{10} = \omega_{10} - \omega_{IDT}$. The color scale shows the normalized coherent emission amplitude from the electrical gate to the IDT. No electrical signal is applied to the IDT. At low applied gate power, the qubit can only be excited at zero detuning, where it emits SAWs in proportion to the applied RF amplitude (constant $|S12|$). At higher power, two photons from the gate at $\omega = \omega_{20}/2$ can excite the $|2\rangle$ state of the qubit, which subsequently emits two phonons. For still higher power, photon-to-phonon conversions of increasingly higher orders are visible. **(B)** Theoretical simulation [(*27*), full quantum model]. The markup shows the values of $\Delta\omega_{10}$ where $n$ photons at frequency $\omega_{0n}/n$ can excite the $n^{th}$ state of the qubit and give rise to the emission of $n$ SAW phonons.

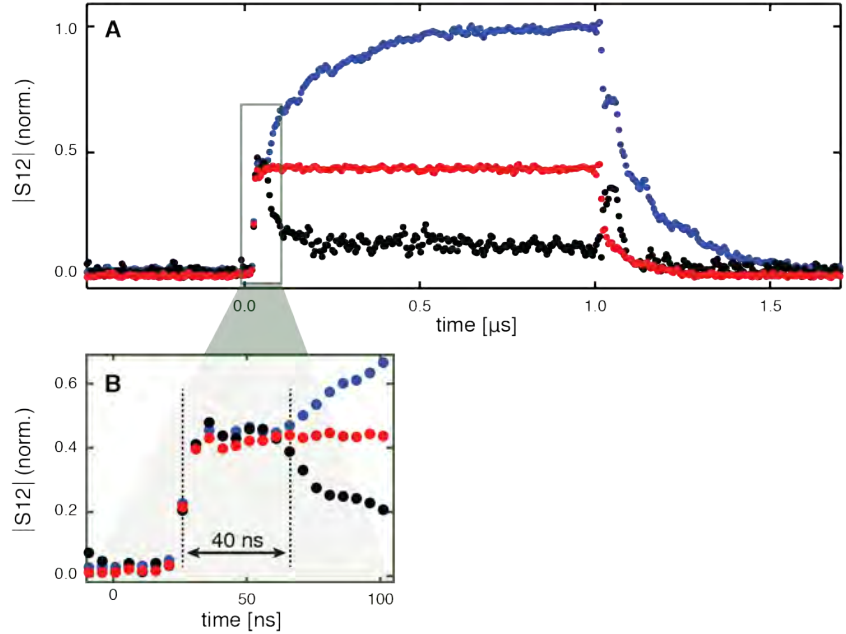

**Fig. 4**. Time-resolved qubit emission. **(A)** By applying microwave pulses to the gate at $\omega = \omega_{IDT}$ and listening to the qubit in the time domain with the IDT, we can clearly determine that the qubit predominantly emits acoustically. When the qubit is strongly detuned ($\omega_{10} \gg \omega_{IDT}$), a certain crosstalk reaches the transmission line of the IDT from the gate (red). This crosstalk is independent of power and frequency, and we attribute it to capacitive coupling between the transmission lines of the gate and the IDT. With the qubit near resonance, $\omega_{10} \approx \omega_{IDT}$, the transduction coefficient increases substantially above the crosstalk level when the acoustic signal adds in phase with the crosstalk (blue) and decreases when the SAW and the crosstalk have opposite phases (black). **(B)** Zoom-in of panel A. The emission from the qubit is observed ~40ns after the onset of the crosstalk, which corresponds to the acoustic propagation time from the qubit to the IDT.

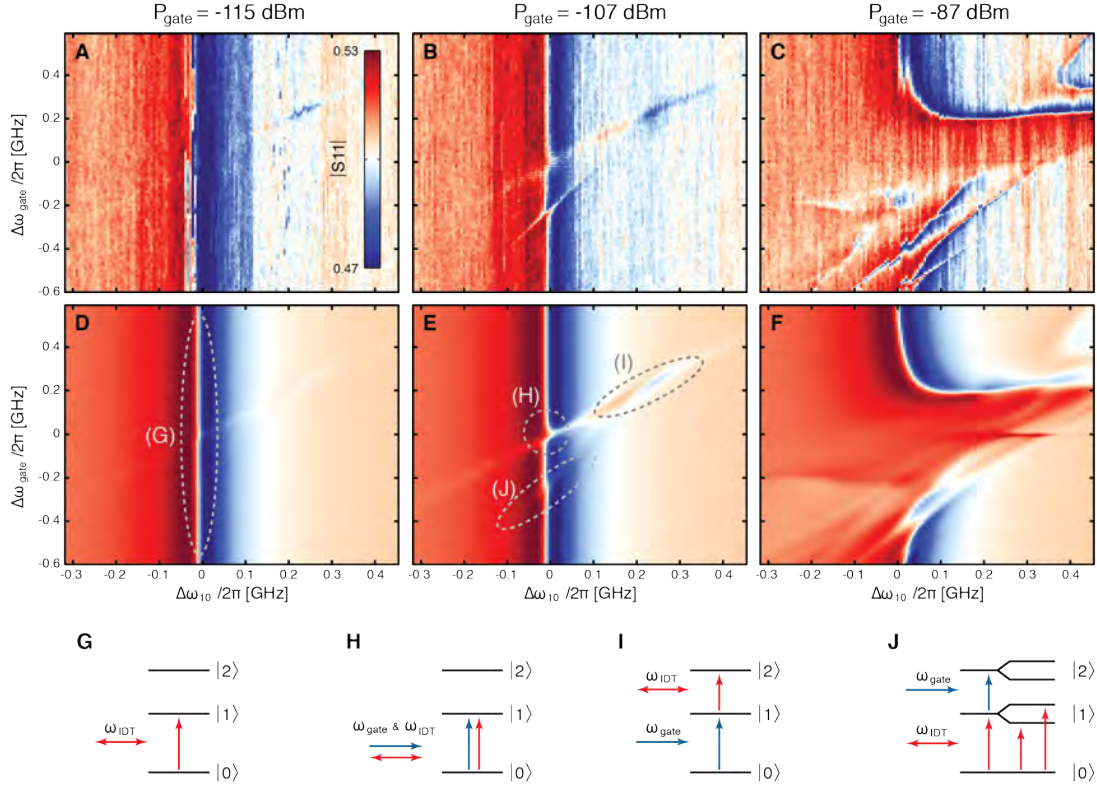

**Fig. 5**. Hybrid two-tone spectroscopy. We apply a continuous acoustic probe tone to the IDT with constant frequency $\omega_{IDT}$ and low power. At the same time, a continuous control tone with varying power $P_{gate}$ and detuning from the probe, $\Delta\omega_{gate} = \omega_{gate} - \omega_{IDT}$, is applied to the gate. The color scale shows the electrical reflection coefficient |S11| of the IDT vs $\Delta\omega_{gate}$ and the qubit detuning $\Delta\omega_{10} = \omega_{10} - \omega_{IDT}$. **(A-C)** Experimental data for increasing $P_{gate}$. **(D-F)** Theoretical simulations corresponding to the experiments in panels A-C. Selected features are highlighted and the corresponding qubit transitions illustrated in panels G-J. **(G)** In the absence of a control signal on the gate, |S11| is modulated by the qubit flux, which is seen in panels A and D as a vertical feature at $\Delta\omega_{10} = 0$. **(H)** For $\Delta\omega_{gate} = \Delta\omega_{10} = 0$, a strong control tone contributes to the saturation of the first qubit transition. **(I)** Provided that the |1⟩ state of the qubit is significantly populated by the control tone, nonlinear acoustic reflection can occur also for $\omega_{21} = \omega_{IDT}$. We observe this when a moderately strong control tone is applied at $\omega_{gate} = \omega_{10}$. **(J)** For moderate to strong electrical driving of the |1⟩ →|2⟩ transition, Rabi splitting of the qubit energy levels gives rise to an Autler-Townes doublet (*35, 36*). For very high control tone power (as shown in panel F), higher-order qubit states are populated and subject to Rabi splitting. This gives rise to a complex set of features which is well captured by the theoretical model [(*27*), full quantum model].

# Supplementary Materials for

## Propagating phonons coupled to an artificial atom

Martin V. Gustafsson[*], Thomas Aref, Anton Frisk Kockum, Maria K. Ekström, Göran Johansson, and Per Delsing[†]
\* correspondence to:  mg3465@columbia.edu
† correspondence to:  per.delsing@chalmers.se

**The supplemetary materials include:**

    Materials and Methods
    Fig. S1.
    Table S1.

# Materials and Methods

## S1. Semi-classical approximation of the qubit

In a common model for an IDT (*22–24*), its electro-acoustic properties are represented by a complex admittance element $Y_a = G_a + iB_a$, which is shunted by the geometric capacitance of the fingers. By applying this model to the finger structure of our qubit and approximating the shunting SQUID as an adjustable inductance $L_J$, we get the semi-classical circuit shown in Fig. S1.

When the finger structure serves as an emitter of SAW, dissipation of electrical power in $G_a$ represents conversion into SAW power, divided equally between the two directions of propagation. On acoustic resonance, $\omega = \omega_{IDT}$, $G_a(\omega)$ is at its maximum and the imaginary element $B_a$ is zero. Off resonance $G_a = G_a(\omega_{IDT})[\sin(x)/x]^2$ with $x = N_{tr}\pi(\omega - \omega_{IDT})/\omega_{IDT}$ and where $B_a$ is the Hilbert transform of $G_a$.

In the following, we limit the discussion to the case of acoustic resonance. Following Datta (*22*), we have

$$G_a(\omega_{IDT}) = -\mu g_m \tag{1}$$

where $\mu$ is the emitter response and $g_m$ the receiver response. $\mu$ relates the amplitude of the emitted SAW to the voltage $V_T$ over $G_a$ as

$$\phi_{em}^{\pm} = \mu V_T \tag{2}$$

with

$$\mu = i c_g K^2 N_{tr}. \tag{3}$$

Here, $N_{tr}$ is the number of finger periods, in our case $N_{tr} = 20$. $c_g$ is a geometry factor that depends on whether the IDT structure has single or double fingers and on the fraction $\eta$ of metallized area. The qubit has double fingers to reduce internal reflections, with $\eta \approx 50\%$. This gives $c_g \approx 0.8$. $K^2$ is a material constant that determines the piezoelectric coupling strength of the SAWs. For Rayleigh waves in the [011] direction on (100)-cut GaAs, $K^2 \approx 7 \times 10^{-4}$.

When a SAW with amplitude $\phi_{in}^+$ impinges on the finger structure, it generates a current $I$ through $G_a$:

$$I = g_m \phi_{in}^+ \tag{4}$$

with

$$g_m = 2\mu \frac{W\epsilon\omega_{IDT}}{K^2}. \tag{5}$$

Here, $W$ is the width of the SAW beam, *ie* the distance over which the fingers from the two electrodes overlap. $\epsilon$ is the effective dielectric constant of the substrate material. For our device, we have $W = 25\,\mu\text{m}$ and $\epsilon = 120\text{ pF/m}$.

The capacitance of the fingers is included as a parallel element to $Y_a$. We have

$$C_{tr} = \sqrt{2}N_{tr}W\epsilon, \tag{6}$$

where the factor $\sqrt{2}$ enters to account for the double fingers.

In addition to $G_a$ and $C_{tr}$, our qubit incorporates a SQUID, shunting the two electrodes as shown in Fig. S1. The SQUID acts as a non-linear inductance, $L_J$, and forms an electrical resonance circuit together with $C_{tr}$. While the acoustic resonance frequency $\omega_{IDT}$ is determined by lithography, the electrical resonance frequency can be tuned by adjusting $L_J$ with a magnetic field. The two resonances coincide for

$$L_J = \frac{1}{\omega_{IDT}^2 C_{tr}}. \tag{7}$$

When this condition is fulfilled, the impedance of the *LC* circuit approaches infinity, and we are left with only the acoustic element $G_a$. As a result, an incoming SAW beam with amplitude $\phi_{in}^+$ produces a voltage over $G_a$ with amplitude

$$V_T = g_m \phi_{in}^+ / G_a, \tag{8}$$

which in turn gives rise to re-radiation of SAW in the forward (+) and backward (-) directions with amplitudes

$$\phi_{em}^\pm = \mu V_T = \frac{\mu g_m}{G_a}\phi_{in}^+ = -\phi_{in}^+. \tag{9}$$

Hence, the net transmission of SAW in the forward direction is $\phi_{in}^+ + \phi_{em}^+ = 0$, and the emission in the backward direction is $\phi_{out}^- = -\phi_{in}^+$. This gives a qualitative picture of the acoustic reflection we observe experimentally. Note that full reflection only ensues when the approximation of the SQUID as a linear inductance is valid, *ie* for low enough SAW power that the qubit is never excited beyond the $|1\rangle$ state. The dependence of the reflection coefficient on power is discussed in S2.

We can also use this model to estimate the coupling $\Gamma_{10}$ of the qubit, which is the rate at which energy stored in the $LC$ resonator converts into SAWs by dissipating in $G_a$. This is given by the damping ratio $\zeta_{RLC}$ of the $RLC$ circuit:

$$\Gamma_{10} = \omega_{IDT}\, \zeta_{RLC} = \omega_{IDT}\, \frac{G_a}{2}\sqrt{\frac{L_J}{C_{tr}}}\ . \tag{10}$$

Using Eqs. (1), (3), and (5) - (7), we get

$$\Gamma_{10} = \frac{\omega_{IDT}\, c_g^2\, K^2 N}{\sqrt{2}} = 2\pi \times 30\,\text{MHz}\ . \tag{11}$$

We find this to be in acceptable agreement with our experimental value of $\Gamma_{10} = 2\pi \times 38$ MHz, considering that $K^2$ is tabulated for room temperature and $\eta$ is inexact for the fine lithographic pitch of the qubit.

It should be noted that the circuit model discussed above does not apply to an IDT with strong internal mechanical reflections, such as the one we use to launch and detect SAWs. The finger structure of the qubit, however, is designed to be free from such reflections.

S2. Full quantum model

The quantum mechanical model of the present experiment is very close to that for recent experiments with a transmon in a 1D transmission line (*4–7*). In both cases, we consider a multi-level (artificial) atom interacting with a continuum of left- and right-moving 1D bosonic fields. The important difference lies in the size of the artificial atom. In previous experiments, the atoms were of negligible size compared to the wavelength of the fields they interacted with, whereas in our case the transmon is several wavelengths long and couples to the fields at several discrete points. We therefore begin this theory overview by showing how a large atom differs from a small one. We then outline the theoretical model used to calculate reflection and transduction coefficients in the various one- and two-tone experiments reported in this article.

The theory presented in this section is a summary of the more detailed work in Ref. (*28*). We consider a multi-level atom with energy levels $|m\rangle$, $m = \{0,1,2,...\}$, coupled to a continuum of left- and right-moving bosonic fields. The fields are described by annihilation operators $a_{Lj}$ and $a_{Rj}$, respectively, with $j$ denoting the phonon mode of frequency $\omega_j$. The atom and the fields interact at $N$ points with coordinates $x_k$, $k = \{1,...,N\}$.

We assume that the time it takes to travel between any two connection points is negligible compared to the relaxation rate of the atom. With this approximation, we only need to take into account a phase factor $e^{i\omega_j x_k / v_0}$ when we write down the interaction Hamiltonian. Under these conditions, we get the Hamiltonian $H = H_A + H_F + H_I$, where we have defined the atom Hamiltonian, the field Hamiltonian, and the interaction Hamiltonian as

$$H_A = \sum_m \hbar \omega_m |m\rangle\langle m|,$$
$$H_F = \sum_j \hbar \omega_j \left(a_{Rj}^\dagger a_{Rj} + a_{Lj}^\dagger a_{Lj}\right), \tag{12}$$
$$H_I = \sum_{j,k,m} g_{jkm} \left(\sigma_+^m + \sigma_-^m\right)\left(a_{Rj} e^{-i\omega_j x_k / v} + a_{Lj} e^{i\omega_j x_k / v} + a_{Rj}^\dagger e^{i\omega_j x_k / v} + a_{Lj}^\dagger e^{-i\omega_j x_k / v}\right).$$

Here, $\sigma_+^m = |m+1\rangle\langle m|$, $\sigma_-^m = |m\rangle\langle m+1|$, and $g_{jkm}$ denotes the coupling strength which can depend on both phonon mode ($j$), connection point ($k$), and atom transition ($m$). We can divide this coupling as $g_{jkm} = g_j g_k g_m$, where $g_j$ can be assumed constant, $g_m = \sqrt{m+1}$, and $g_k$ is a factor describing the ratio between coupling strengths at different points.

Carrying through the derivation of a master equation by tracing out the bosonic modes and making the rotating wave approximation (RWA), we see that the crucial difference compared to the well-known case of a small atom (*29*) is that the constant coupling $g_j$ is replaced with a frequency dependent factor

$$A(\omega_j) = g_j \sum_k g_k e^{-i\omega_j x_k / v_0} \qquad (13)$$

in the interaction Hamiltonian. The result is that we get a master equation

$$\dot{\rho}(t) = -\frac{i}{\hbar}\left[\sum_m \hbar(\omega_m + \Delta_m)|m\rangle\langle m|, \rho(t)\right] + \sum_m \Gamma_{m+1,m} \mathcal{D}[\sigma_-^m], \qquad (14)$$

where we have introduced the Lindblad superoperator $\mathcal{D}[X]\rho = X\rho X^\dagger - \frac{1}{2}X^\dagger X \rho - \frac{1}{2}\rho X^\dagger X$. The relaxation rates are given by

$$\Gamma_{m+1,m} = 4\pi g_m^2 J(\omega_{m+1,m})|A(\omega_{m+1,m})|^2 \qquad (15)$$

and the Lamb shifts are

$$\Delta_m = 2\mathcal{P}\int_0^\infty d\omega \frac{J(\omega)}{\omega}|A(\omega)|^2 \left(\frac{g_m^2 \omega_{m+1,m}}{\omega + \omega_{m+1,m}} - \frac{g_{m-1}^2 \omega_{m,m-1}}{\omega - \omega_{m,m-1}}\right). \qquad (16)$$

Here, we have assumed zero temperature and we have defined a density of states $J(\omega)$ for the bosonic modes. The difference compared to the case of a small atom is the appearance of the factor $|A(\omega)|^2$ instead of merely $g_j^2$ in the two expressions above, giving a frequency dependence for the relaxation rates and for the Lamb shift.

In the present experiment, we can assume that the coupling strengths are equal at each connection point, *ie* $g_k = 1$ for all $k$, and that the spacing between subsequent coupling points is equal. We can then define a phase $\phi = \omega(x_{k+1} - x_k)/v_0$ and evaluate the factor $A(\omega)$ under these constraints, which gives a frequency dependence for both the relaxation rates and the main contribution to the Lamb shift

$$\Gamma_{m+1,m} = \Gamma_{m+1,m}^{\text{small}} \frac{\sin^2\left(\frac{N}{2}\phi\right)}{\sin^2\left(\frac{1}{2}\phi\right)},$$

$$\Delta_m = \Gamma_{m,m-1}^{\text{small}} \frac{N\sin(\phi) - \sin(N\phi)}{2(1-\cos(\phi))}, \qquad (17)$$

where $\Gamma_{m,m+1}^{\text{small}}$ denotes the relaxation rate that would result for a small atom which has a single coupling point with $g_{k=1} = 1$. The frequency dependence of the relaxation rate and the Lamb shift for the first transmon transition coincides with what one gets from a classical calculation for an IDT shunted by an inductance such that its resonance frequency coincides with that of the transmon, as described in S1.

In the present experiment, the effect of the Lamb shift is small and we can neglect it in the simulations. Since we only probe the qubit acoustically around a single frequency ($\omega_p \approx \omega_{IDT}$) in all experiments, the frequency dependence of the relaxation rate is also not prominent. However, in future experiments using a wider range of phonon frequencies, the effects described in this section will certainly be interesting to investigate.

To model the experiment where an acoustic probe of varying strength is reflected from the transmon, we use the theory developed for a two-level atom in an open 1D transmission line, see *e.g.* Ref. (*31*). We assume that the transmon relaxes primarily through phonons, equally divided between the forward and backward directions, and neglect all other relaxation channels. In the rotating frame of the probe frequency $\omega_p$, the Hamiltonian for the transmon is

$$H = \frac{\Delta\omega_{10}}{2}\sigma_z - i\frac{\Omega}{2}(\sigma_+ - \sigma_-), \tag{18}$$

where $\Delta\omega_{10} = \omega_{10} - \omega_p$ is the detuning and $\Omega = \alpha\sqrt{2\Gamma_{10}}$ is the Rabi frequency determined by the total phonon relaxation rate $\Gamma_{10}$ and the incoming phonon flux of $|\alpha|^2$ phonons per second. The master equation for our open quantum system gives the Bloch equations

$$\partial_t \langle\sigma_-\rangle = (-i\Delta\omega_{10} - \gamma)\langle\sigma_-\rangle + \frac{\Omega}{2}\langle\sigma_z\rangle,$$
$$\partial_t \langle\sigma_+\rangle = (i\Delta\omega_{10} - \gamma)\langle\sigma_+\rangle + \frac{\Omega}{2}\langle\sigma_z\rangle, \tag{19}$$
$$\partial_t \langle\sigma_z\rangle = -\Gamma_{10}(1 + \langle\sigma_z\rangle) - \Omega(\langle\sigma_-\rangle + \langle\sigma_+\rangle),$$

where $\gamma = \Gamma_\phi + \frac{1}{2}\Gamma_{10}$ is the decoherence rate. In our system, we expect the dephasing rate $\Gamma_\phi$ to be low due to the high $E_J/E_C$ ratio. Our numerical implementations of the theoretical model take dephasing into consideration. Since we find it to be negligible, we exclude it from the following treatment in the interest of transparency.

From the steady-state solution of the Bloch equations, we then find the reflection coefficient

$$r = \sqrt{\Gamma_{10}/2}\langle\sigma_-\rangle/\alpha = -\frac{1 - 2i\dfrac{\Delta\omega_{10}}{\Gamma_{10}}}{1 + 4\left(\dfrac{\Delta\omega_{10}}{\Gamma_{10}}\right)^2 + 2\left(\dfrac{\Omega}{\Gamma_{10}}\right)^2}. \tag{20}$$

When driving the transmon via the electric gate and listening to the outgoing phonons, we can no longer neglect the small coupling that allows the transmon to also relax to photons via the electrical gate. Defining the relaxation rate to phonons $\Gamma_{ac}$ and the relaxation to photons $\Gamma_{el}$, we again use the steady-state solutions of the Bloch equations (19) with $\Gamma_{10} = \Gamma_{ac} + \Gamma_{el}$ and $\Omega = 2\alpha\sqrt{\Gamma_{el}}$. The resulting transduction coefficient is

$$t = \frac{\sqrt{\frac{\Gamma_{ac}}{2}}\langle\sigma_-\rangle}{\alpha} = \sqrt{\frac{\Gamma_{ac}\Gamma_{el}}{2}} \frac{i\Delta\omega_{10} - \frac{\Gamma_{10}}{2}}{\Delta\omega_{10}^2 + \frac{\Gamma_{10}^2}{4} + \frac{\Omega^2}{2}}, \quad (21)$$

again assuming no dephasing. Here, the detuning is $\Delta\omega_{10} = \omega_{10} - \omega_{gate}$, where $\omega_{gate}$ is the frequency of the electric drive.

To fully model the transduction experiment, we need to include more levels in the transmon. We do this by numerically solving the master equation for $N_t$ transmon levels,

$$\dot{\rho}(t) = -\frac{i}{\hbar}\left[\sum_{m=0}^{N_t-1}\hbar\Delta\omega_m|m\rangle\langle m| - i\frac{\Omega}{2}(\Sigma_+ - \Sigma_-), \rho(t)\right] + \sum_{m=0}^{N_t-1}\Gamma_{m+1,m}\mathcal{D}[\sigma_-^m], \quad (22)$$

where we have defined

$$\Sigma_- = \sum_{m=0}^{N_t-1}\sqrt{m+1}|m\rangle\langle m+1|, \quad (23)$$

$\Sigma_+ = \Sigma_-^\dagger$, and where $\Gamma_{m+1,m} = (m+1)\Gamma_{10}$. We work in the rotating frame of the electric drive frequency $\omega_{gate}$, where $\Delta\omega_m = \omega_m - m\omega_{gate}$. The transduction coefficient is given by

$$t = \frac{\sqrt{\frac{\Gamma_{ac}}{2}}\langle\Sigma_-\rangle}{\alpha}, \quad (24)$$

where the expectation value is computed from the steady-state solution of the master equation.

In the hybrid two-tone spectroscopy experiment, we drive the transmon at a frequency $\omega_{gate}$ through the electric gate and monitor the reflection of a weak acoustic probe at frequency $\omega_p$. To find the reflection coefficient for the weak acoustic probe, we follow (*32*) to numerically compute the susceptibility

$$\chi(\omega_p) = i\int_0^\infty dt\, e^{i\omega_p t}\langle[\Sigma_-(t), \Sigma_p(0)]\rangle, \quad (25)$$

where $\Sigma_p = -i(\Sigma_+ - \Sigma_-)$. The expectation values are given by (29)

$$\langle \Sigma_-(t)\Sigma_p(0) \rangle = \text{Tr}\left(\Sigma_- e^{-i\omega_{gate}t} e^{\mathcal{L}t} \Sigma_p \rho_{ss}\right), \tag{26}$$

$$\langle \Sigma_p(0)\Sigma_-(t) \rangle = \text{Tr}\left(\Sigma_- e^{-i\omega_{gate}t} e^{\mathcal{L}t} \rho_{ss} \Sigma_p\right), \tag{27}$$

where $\mathcal{L}$ is the Liouvillian, defined by $\dot{\rho} = \mathcal{L}\rho$, and $\rho_{ss}$ is the steady-state solution of Eq. (22) (we average over different phases of the electrical drive). Considering the weak probe of strength $\alpha_p$ to give a perturbation to the Hamiltonian,

$$H_{probe} = \sqrt{\Gamma_{ac}/2}\, \alpha_p \Sigma_p, \tag{28}$$

Kubo theory (33) for linear response tells us that the susceptibility lets us connect $\langle \Sigma_- \rangle$ to the susceptibility through

$$\langle \Sigma_- \rangle = \chi(\omega_p)\sqrt{\frac{\Gamma_{ac}}{2}}\alpha_p, \tag{29}$$

and we get the reflection coefficient

$$r = \frac{\Gamma_{ac}}{2}\chi(\omega_p). \tag{30}$$

S3. Data analysis and parameter fitting

In Figs. 2B and 2D-F, we have subtracted the complex reflection acquired with the qubit detuned. These figures thus show the change in reflection coefficient due to the qubit alone. In Fig. 3A, we have subtracted the background transmission coefficient, acquired with the qubit detuned. In Fig. 5A-C, we do not subtract any background, since this would decrease the contrast of the finest features. The simulations in Fig. 5D-F are processed to take the effects of interference and on-chip acoustic reflections into account. The levels of the data sets have been adjusted for internal consistency, to compensate for variations in the amplifier gain.

We determine the flux periodicity of the qubit from the resonance points in Fig. 2B. From the positions of the transduction peaks in Fig. 3A we find the higher energy levels of the qubit and can thus extract $E_J$ and $E_C$ to good precision. The value we find for $E_C$ agrees well with the value for $C_{tr}$ that we calculate based on the qubit geometry.

S2 gives analytical expressions for the acoustic reflection coefficient $r$ and electro-acoustic transduction coefficient $t$ of the qubit [Eqs. (20) and (21), respectively]. When we use these formulas to characterize our device, we must also account for losses and on-chip reflections of electrical and acoustic signals, defined by the scattering parameters $S11_0$, $S12_0$, $S22_0$ and $\theta_L$.

$S11_0$ is the electrical reflection coefficient of the IDT in the absence of reflections from the qubit. $S12_0$ is the net conversion at $\omega = \omega_{IDT}$ between an electrical signal applied to the IDT and SAW reaching the qubit. $S22_0$ is the acoustic reflection coefficient of the IDT, seen from the qubit. $\theta_L$ is the phase shift acquired by the SAW upon propagation between the IDT and the qubit. In addition, we used as fitting parameters the coupling rate $\Gamma_{10}$ of the qubit to SAWs and the (much lower) coupling rate $\Gamma_{gate}$ between the qubit and the gate.

With multiple reflections between the qubit and the IDT accounted for, these parameters combine as follows to give the total electrical reflection coefficient for a steady state signal applied to the IDT:

$$S11 = S11_0 + \frac{S12_0^2}{S22_0} \sum_{n=1}^{\infty} (r\, S22_0\, e^{2i\theta_L})^n = S11_0 + \frac{S12_0^2}{\frac{e^{-2i\theta_L}}{r} - S22_0}. \qquad (31)$$

Here, $r$ is the acoustic reflection coefficient of the bare qubit, given by Eq. (20) in S2.

$S11_0$ is readily extracted from the data shown in Fig. 2A, and we determine $S22_0$ from the stepwise change of signals acquired in the time domain (Fig. 4). The remaining parameters are adjusted for optimum agreement between experimental data and theory, as shown in Figs. 2E, 2F, 3, and 5. The acoustic coupling rate $\Gamma_{10}$ is of particular interest, and its value corresponds to the width of the peak in Fig. 2E. The value we find is consistent with the high-power saturation of the acoustic reflection, provided that we assume a moderate change in the attenuation of the microwave components between ambient and cryogenic conditions. We include the pure dephasing rate $\Gamma_\phi$ in the fitting procedure, but find that it is negligible compared to $\Gamma_{10}$ and that the best fitting is achieved with $\Gamma_\phi = 0$.

In addition to these data sets, the optimized parameters produce theoretical agreement for the dependence on flux and applied power of gate signals transduced by the qubit into SAWs directed toward the IDT.

All parameters extracted from the experimental data are shown in Table S1.

S4. Sample preparation

We fabricated the sample on the polished (100) surface of a semi-insulating GaAs wafer, with the propagation direction of the SAW aligned with the [011] direction of the crystal. All lithographic patterns were defined by electron beam lithography with 100 keV electron energy. Each metal layer was deposited by evaporation through a resist mask followed by lift-off in a solvent.

The first layer, made of gold with a thin underlying sticking layer of titanium, consists of alignment marks and labels, and all subsequent patterns were aligned to this initial layer. This layer is not shown in Fig. 1.

The second set of metal features includes the finger structures of the IDT and the qubit as well as the electrical gate. The layer stack consists of 27 nm of aluminum capped with 3 nm of palladium, shown in blue in Figs. 1A-E. The palladium prevents oxidation of the aluminum and allows subsequent metallization to make contact with the finger structures. These structures extended over more than 50 μm, yet the line widths and spacings of the qubit fingers is less than 100 nm, which makes the lithography sensitive to electron backscattering. To compensate for this, we applied a numerical dose correction and rescaled the pattern to counteract broadening of the lines.

Next, we deposited the large scale ground planes and transmission lines that are shown in yellow in Figs. 1A and 1B. These consist of 5 nm of titanium, 85 nm of gold, and 10 nm of palladium. The transmission lines leading from the edges of the chip to the IDT and the gate have a nominal characteristic impedance of 50 Ω. The ground planes were designed to provide electrical screening between the IDT and the qubit without impeding SAW propagation.

In the final lithographic step, we defined the SQUID of the qubit, shown in green in Figs. 1A, 1D, and 1E. We deposited the SQUID by two-angle evaporation of aluminum with an intermediate step of in-situ oxidation in pure $O_2$ gas. The palladium cap on the finger structure allows us to make reliable contact between the SQUID and the rest of the qubit. Since the cap layer is thin, it becomes superconducting at cryogenic temperatures along with the aluminum by the proximity effect.

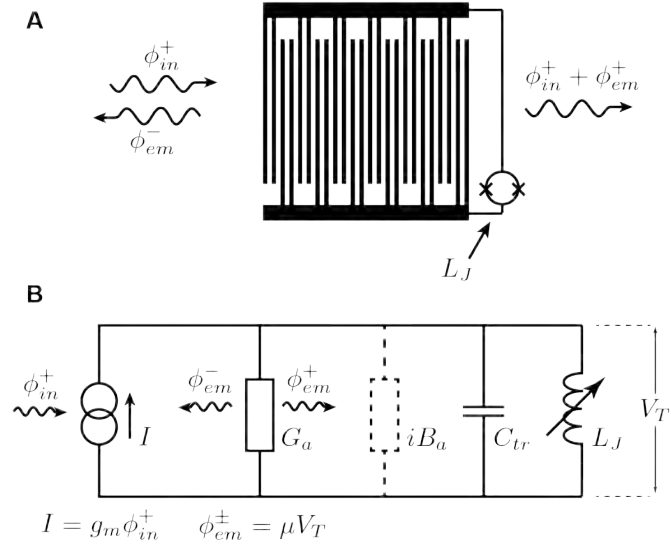

**Fig. S1.**
Semi-classical model for the acoustically coupled qubit. **(A)** Layout of the qubit, with incoming and outgoing SAW components shown. The capacitively coupled gate is not included in this model. **(B)** Equivalent circuit, see text.

**Table S1.**

Sample parameters extracted from measured data. The scattering parameters are given in amplitude units.

| Parameter | Value |
|---|---|
| $S11_0$ | 0.51 |
| $S12_0$ | 0.28 |
| $S22_0$ | 0.55 |
| $\theta_L$ | 0.61 rad |
| $\Gamma_{10}$ | $2\pi \times 38$ MHz |
| $\Gamma_{el}$ | $2\pi \times 750$ MHz |
| $\Gamma_\phi$ | $< 0.1\, \Gamma_{10}$ |